# Degradation Detection Method for Railway Point Machines

Chong Bian[1], Shunkun Yang*, [2], Tingting Huang[2], Qingyang Xu[3], Jie Liu[2], Enrico Zio[4, 5, 6]

1.School of Automation Science and Electrical Engineering, Beijing University of Aeronautics and Astronautics, Beijing 100191, China

2.School of Reliability and Systems Engineering, Beijing University of Aeronautics and Astronautics, Beijing 100191, China

3.Infrastructure Inspection Research Institute, China Academy of Railway Sciences, Beijing 100191, China

4.Chair on system science and energetic challenges, EDF Foundation, Laboratoire Genie Industriel, CentraleSupélec, Université Paris-Saclay, 3 rue Joliot Curie, 91190 Gif-sur-Yvette, France

5.Energy Department, Politecnico di Milano, Piazza Leonardo da Vinci, 32, 20133 Milano, Italy

6.Sino-French Risk Science and Engineering Lab, Beijing University of Aeronautics and Astronautics, Beijing 100191, China

## Abstract

Point machines (PMs) are used for switching and locking railway turnouts, and are considered one of the most critical elements of a railway signal system. The failure of the point mechanism directly affects the operation of the railway and may cause serious safety accidents. Hence, there is a need for early detection of the anomalies in PMs. From normal operation to complete failure, the machine usually undergoes a series of degradation states. If the degradation states are detected in time, maintenance can be organized in advance to prevent the malfunction. This paper presents a degradation detection method that can effectively mine and identify the degradation state of the PM. First, power data is processed to obtain the feature set that can describe the PM characteristics effectively. Then, a clustering analysis of the feature set is carried out by self-organizing feature-mapping network, and various degradation states are mined. Finally, the optimized support vector machine is used to build the state classifier to identify the degradation state of the PM. The experimental results obtained with the Siemens S700K PM show that the proposed method could not only mine the effective degradation states, but also obtain high identification accuracy.

## 1. Introduction

In the last decade, the construction of the China Railway High-speed has made a qualitative leap, and the railway network is moving from "four vertical and four horizontal" to "eight vertical and eight horizontal." As of 2017, the total mileage of China railway has exceeded 124,000 kilometers, of which the high-speed rail mileage was 22,000 kilometers, accounting for 60% of the total mileage of the world. The rapid development of high-speed railway not only shortens the travel time for passengers, but also promotes the coordinated development of the regional economy. Meanwhile, the reliable operation of the train becomes increasingly important. PMs, which change the direction of the train by carrying out switching and locking functions at turnouts, are electromechanical equipment in the railway signal system.

The performance of a PM directly affects the safety of train operation. In the field investigation of the Chenzhou West Station, the related faults caused by the PM were found to account for 30% of all signal equipment failures, and the average time for repairing the PM is approximately 23.4 minutes, which has a serious impact on the operation of the high-speed railway. Therefore, it is necessary to detect the faults in the PM in time and make reasonable maintenance plans to improve the operational efficiency of the high-speed railway.

A condition monitoring system (CMS) can measure, centralize, and analyze the data from sensors installed in the field equipment to detect equipment failure [1]. Many railway companies have implemented CMS for PMs. However, experience in this field has shown that these systems create an alarm only when a failure occurs or very close to the point of failure. In addition, most of those currently in use do not effectively detect faults, and the system simply creates an alert to indicate abnormal behavior, leading to a delay in repairing of the PM [2]. The fault detection technology of CMS needs to be improved in order to increase the maintenance efficiency of PMs. In recent years, research on fault detection for PMs has gradually transitioned from the threshold-based method to statistical analysis, model-based, and classification methods [3]. Compared to thresholding technology, these three methods can ensure early detection of machine failure. Among them, the classification method is widely used in CMS [4].

The classification method can be divided into three main steps, viz. data feature processing, classifier establishment, and fault identification. The data feature processing step converts the measurement signal of the PM, e.g. power signal, current signal, and sound signal, into a feature data set, which can effectively characterize the machine. The classifier establishment is for applying an appropriate intelligent algorithm to learn the feature set and obtain a model that can effectively identify the machine features. The fault identification step employs a classifier to analyze the features of signal data, detect machine failure, and determine the specific fault type. In the classifier establishment step, fuzzy logic, neural network, and support vector machine (SVM) are commonly used [5-7]. SVM is a margin-based classifier based on small sample learning with good generalization capabilities, which is frequently used in classification applications [8]. Asada et al. [9] proposed a condition-monitoring method combining wavelet transform (WT) and SVM. SVMs were used to learn the power-signal feature data extracted by WT and the different fault conditions of the PM were identified on the basis of the SVM model. Vileiniskis et al. [4] proposed a fault-detection method based on a one-class SVM (OCSVM). By preprocessing the current signal to obtain specific feature data and using these data to train the OCSVM, the obtained model was employed to detect the possible fault of the PM in advance. Lee et al. [3] proposed a fault-detection and diagnosis method based on mel-frequency cepstral coefficients (MFCC) and SVM. The selected MFCC feature data of the sound signal was used to train the SVMs, which were employed to detect and classify the fault sound of the PM. Although these classification methods can accurately identify different types of fault states, they cannot detect the degradation state of the PM. The degradation state is a state between the normal and fault states of a machine, which will gradually deteriorate with the passage of time and eventually change to a fault state [10]. Although degradation is not the root cause of PM damage, it increases the probability of failure and makes maintenance difficult.

Recently, numerous studies have been conducted on the degradation of electrical machines, with data-driven methods, which mine the hidden information of the monitoring data through data analysis technology to measure and detect the degradation of the machine, being widely used [11-15]. Data-driven methods are generally divided into two categories: artificial intelligence (AI) and statistical methods [16]. A neural network is the most commonly used AI method, and because of its powerful learning ability, it acts as a recognizer for distinguishing the degradation state of the machine. Self-organizing feature mapping (SOM) is an unsupervised neural network, and owing to its topology-preserving property, input data with similar features can be mapped to nearby neuron regions in the output space. In degradation studies, SOM is used for anomaly detection and cluster analysis. Yu [17] used normal samples to train SOM and calculated the degradation indicators of monitoring data based on this baseline model, by comparing the difference between the indicators to identify whether the machine is undergoing degradation. German et al. [18] trained SOM using fault samples and identified the degradation of the machine by analyzing the regions in the monitoring data mapped by a model. Hu et al. [19] used SOM to cluster the monitoring data and determined the change trajectory of machine degradation by analyzing the mapping results. However, these methods can only indicate whether the machine is undergoing degradation, and cannot truly identify the specific degradation state. If the type of degradation state can be detected in time during the operation of the machine, the maintenance can be organized in a targeted way to prevent the occurrence of failure effectively.

To solve the above problems, this work presents a degradation detection method based on SOM and SVM, with the Siemens S700K alternating current (AC) PM, which is widely used in the China high-speed railway, as the object of study. In this method, feature processing is first performed on the PM power signal to obtain a feature set that effectively describes the characteristics of the machine. The clustering analysis of the feature set is carried out by SOM to mine various degradation states. Using the optimized SVM to build the classifier, the identification of the degradation state of the PM is realized. The merits of the proposed method are summarized as follows: (1) It can mine the degradation state of the PM with different characteristics. (2) It builds a multi-class classifier that can effectively identify the different degradation states of the PM.

The rest of this paper is organized as follows. Section 2 introduces the background of the PM. The proposed method is described in Section 3. The experimental results are shown in Section 4. Section 5 draws the conclusions.

## 2. Railway point machine monitor

### 2.1 Turnout and point machine

A turnout is an important part of a railway signal interlocking system, and its main function is to change tracks by moving the rails before the train passes, as shown in Fig. 1. The turnout system consists of two parts, indoor and outdoor. The indoor part contains the interlocking console and control circuits. When the interlocking console issues an instruction, the instruction is transmitted to the outdoor mechanical components via the control circuits. According to the instruction, the components change the position of the turnout rail from normal to reverse or vice versa. The outdoor part, as shown in Fig. 2, is mainly composed of a switch, cross guard rail, and lead section. The switch is used to change the turnout position, including the PM, tongue rail, stock rail, point rail, etc. The cross guard rail is used to ensure the safety of the train wheels passing through the intersection of the two tracks. The lead section is used to connect the switch and the cross guard rail. The PM is the actuator of the turnout system, as shown in Fig. 3, and it provides traction for the to-and-fro movement of the tongue rail to complete the conversion of the turnout position. Once the PM fails, it will directly affect the normal operation of the turnout system. Therefore, it is necessary to monitor the running condition of the field PM and detect its abnormality before the failure in order to avoid the occurrence of safety accidents.

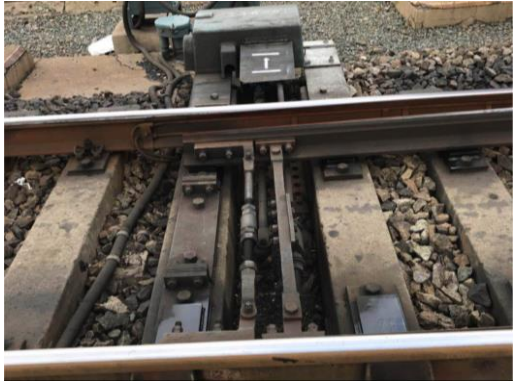
Fig. 1. Tracks moved by the turnout

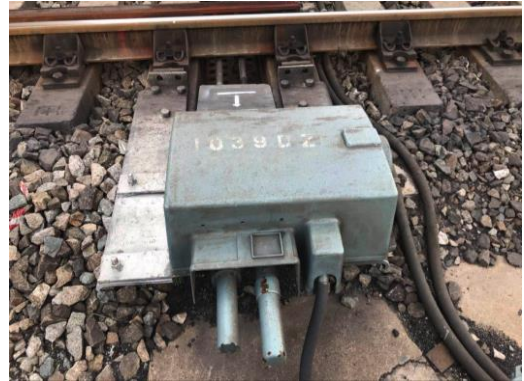
Fig. 3. Railway point machine

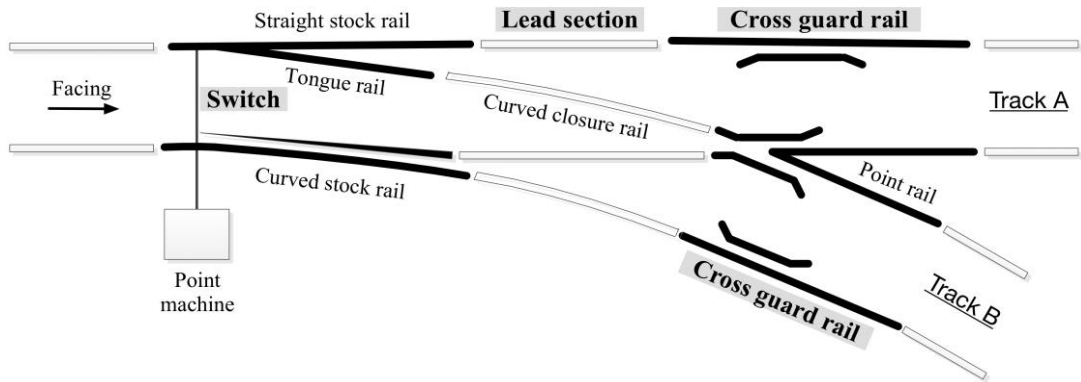
Fig. 2. Outdoor part of the turnout system

### 2.2 Centralized signaling monitoring system

The centralized signaling monitoring system (CSM) is a comprehensive monitoring platform for high-speed railway signal equipment. As one of the CMSs, this system can monitor and record the running data of the equipment in real time, providing a basis for railway departments to grasp the current status of the equipment and carry out maintenance work. CSM can monitor the operation of the PM according to the condition of the turnout relay. CSM mainly monitors the power, current, and voltage data of the PM during the turnout position change process. The real-time power of an alternating current (AC) -PM can be calculated by using the formula $p = U \cdot I \cdot \cos\theta$, where $\theta$ is the angle between phase voltage $U$ and phase current $I$.

During the operation of the PM, the peak value of $U$ is essentially constant, and the peak value of $I$ only changes significantly in the start and end phases. It is $\theta$ that really affects the change in power value. The process of CSM acquiring power data of the PM is shown in Fig. 4. The switch-amount acquisition module obtains the start and end times of the outdoor PM by judging the state of turnout relay 1DQJ, and the indoor devices acquire data during this period. The Hall current sensor collects the current data of each phase at the 21st, 41st, and 61st nodes of the open-phase protector, and the output terminals of the Hall sensor are connected to the power collector. The power collector collects the voltage data of each phase at the 11th, 31st, and 51st nodes of the open-phase protector. In addition, the sampling frequency of the current and voltage is 25 Hz. After the power data is obtained by calculating the current and voltage data, the power collector transmits the power data to the communication front-end processor via the RS-485 bus and finally sends the data to the monitoring host through the switch.

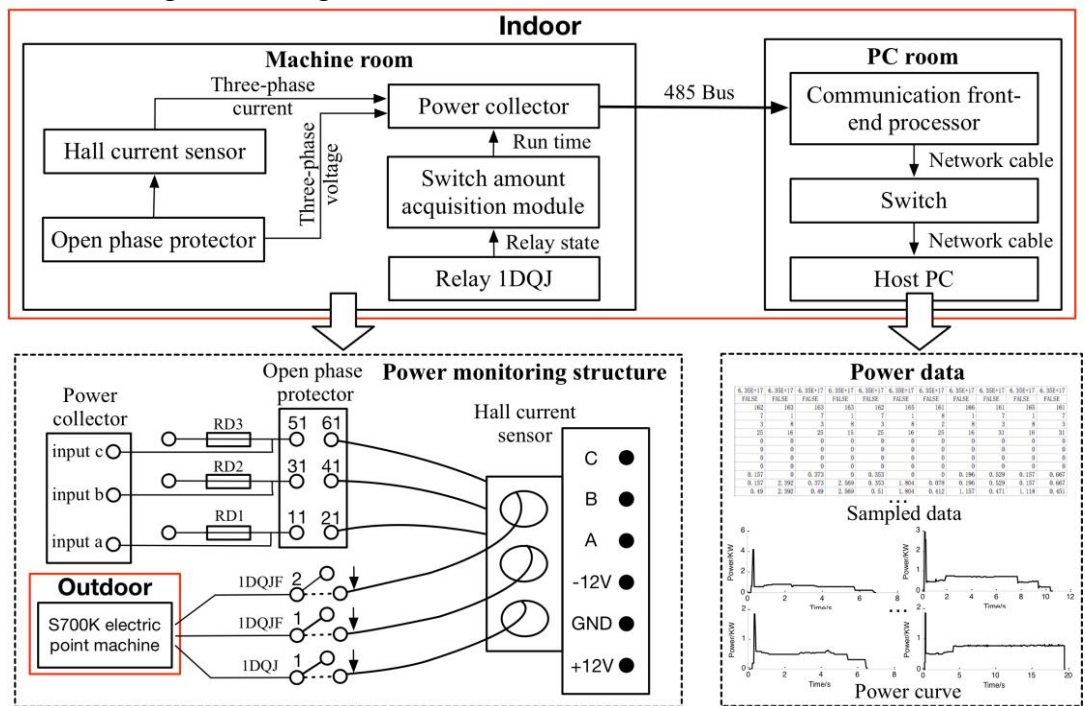

Fig. 4. Flowchart of point-machine power acquisition

For the S700K PM, it takes approximately 6.6 s to complete the conversion of the turnout position normally. Since the sampling interval of the power data by the CSM is 40 ms, the number of data points acquired during the PM operation is about 165. The normal operating power curve of the PM is shown in Fig. 5. As the power source of the turnout, the change in the PM curve is closely related to the former. Therefore, according to the characteristics of the turnout operation, we divide the power curve into four different phases, viz. start-release phase (0–1 s), switch phase (1–4 s), lock phase (4–5 s), and indication phase (≥5 s). (1) In the start-release phase, when the turnout relay 1DQJ is energized, the 1DQJ circuit is connected. The CSM begins to acquire power data of the PM. Because of the time-delay effect of the PM motor circuit, the power values of the curve in the starting region are 0. Subsequently, the motor circuit is connected, and the PM begins to release the turnout. During the releasing process, the PM needs to overcome the large resistance to drive the transmission device, so it will produce a peak value, which is between 1 kW and 2 kW. After the release, the power value of the PM immediately drops and then tends to be stable. The duration of this phase is approximately 1 s. (2) In the switch phase, the PM uses the driving rod to pull the tongue rail to move until it is in close contact with the stock rail. Meanwhile, the detector rod of the PM moves to the designated locking position. During the switching process, the power value of each point is distributed around 0.5 kW because of the smooth operation of the PM. The duration of this phase is approximately 3 s. (3) In the lock phase, the detector rod triggers the locking device of the PM to fix the tongue rail and point rail. The duration of this phase is approximately 1 s. (4) In the indication phase, the motor circuit is cut off and the PM stops running. Because of the slow-release characteristic of relay 1DQJ, it is not released synchronously with the motor circuit cutting off, so the CSM still acquires the PM data. Therefore, a slow-release region is generated at this phase, and the power value of each point in the region fluctuates around 0.2 kW. After the relay is released, its circuit is cut off and the CSM stops acquiring data from the PM. Finally, the indication circuit is connected to determine whether the turnout is normally converted to the specified position. The duration of this phase is approximately 1.6 s. It can be seen that the power data acquired by CSM can specifically describe the changes in electrical parameters and forces during the operation of the PM and accurately reflect its working conditions. Thus, we use the power data to detect the degradation of the PM in this work.

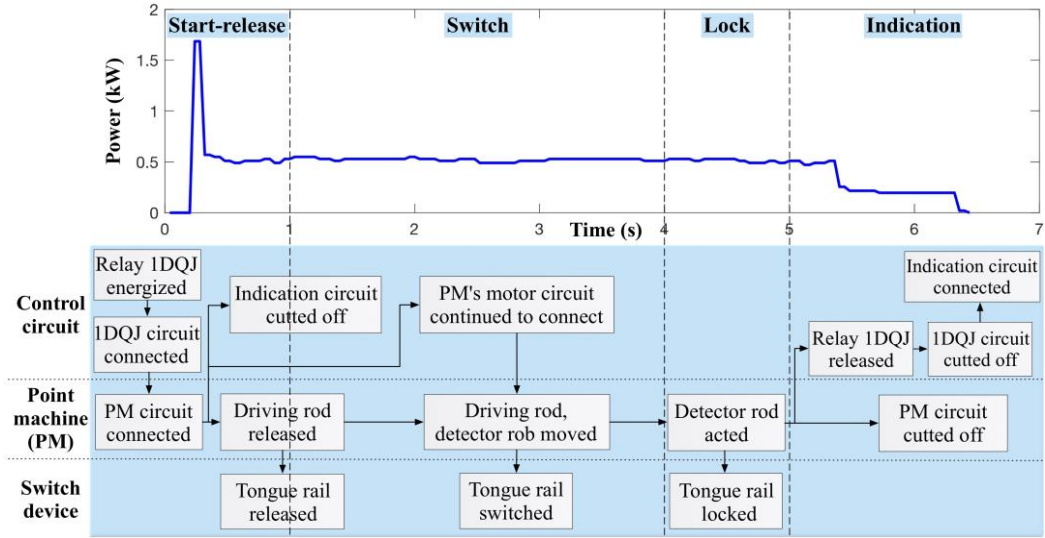

Fig. 5. Normal power curve of the S700K point machine

# 3. Methodology

In this paper, a combined method based on SOM and SVM is proposed to detect the degradation of the PM. This method mainly consists of the following four parts, as shown in Fig. 6: (1) Data acquisition. Use the CSM to acquire the power data of the on-site PM to obtain the sample data set. (2) Feature processing. Perform feature extraction, selection, and further dimensionality reduction on the data set to obtain a low-dimensional feature set. (3) State mining. Based on SOM, clustering analysis of the non-fault feature set is conducted to obtain degradation data with different characteristics. (4) State identification. Build the degradation state classification model based on SVM, in which the particle swarm optimization (PSO) algorithm is used to obtain the optimal parameters of SVM. The obtained model can accurately identify the degradation state of the PM. In the proposed method, SOM and SVM are used to mine and identify the degradation data, respectively. In Section 2, we have shown the acquisition process of the power data and analyzed the characteristics of the power curve in detail, so the last three parts of the proposed method will be studied in this section.

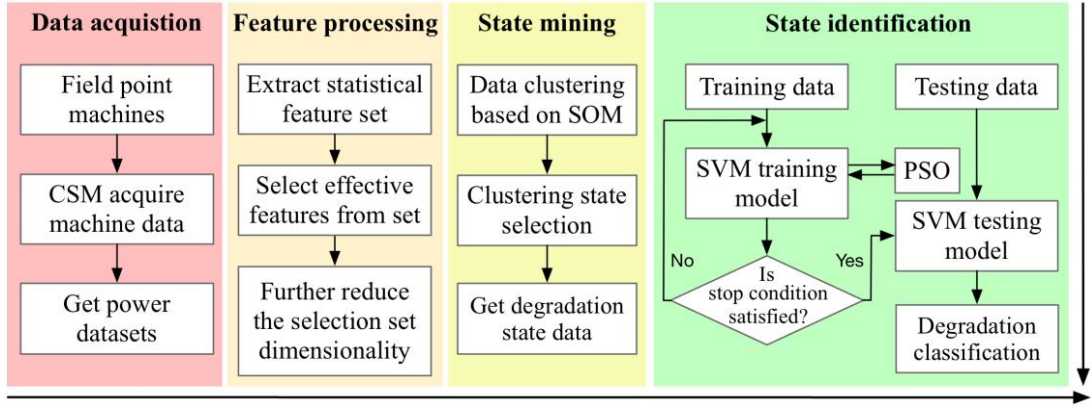

Fig. 6. Framework of the proposed method

## 3.1 Feature processing for power data

### *3.1.1 Statistical feature extraction*

In the process of converting turnout position by the PM, about 165 power data points are acquired by the CSM. Besides, if the machine operates in an abnormal condition, e.g., excessive convert resistance, motor idling, its run time will be prolonged, resulting in the system collecting data points much more than the above number. If we use SOM and SVM to analyze these high-dimensional power data directly, it will generate huge calculations and consume unnecessary time. Hence, statistical features from the power data that can represent the operating state of the PM need to be extracted. In section 2.2, the power curve was divided into four phases. On this basis, 10 parameters of power data in each phase, viz. out-to-in value, maximum difference, mean value, root mean square, variance, sum of difference, kurtosis, crest factor, form factor, and impulse factor, are extracted respectively as the time domain statistical features of the PM, and these parameters are used in [20, 21]. However, it is not sufficient to extract the feature data of the PM only from the time domain. The distribution of the power value directly reflects the change in PM traction. For the fault state related to the traction, the symptom features are concentrated in the specific power value segment. If we do not analyze the power data from the value domain, important features describing the state of the PM will be omitted. According to the operating characteristics of the PM, we divide the power value into three segments, viz. slow-release segment (0–0.3 kW), switch segment (0.3–0.7 kW), and release segment ($\geq 0.7$ kW). Further, we project the power curve onto the value domain and extract eight parameters, viz. maximum time value, mean value, number of data points, maximum difference, median, maximum value, time median, and mode, in each segment as the value domain statistical features to describe the operating states of the PM more completely. For a power sample acquired by the CSM, its 64-dimensional feature vector $x$ can be expressed as:

$$x = (t_{1,1}, \cdots, t_{i,j}, \cdots, t_{4,10}, v_{1,1}, \cdots, v_{m,n}, \cdots, v_{3,8}) \ . \tag{1}$$

where $t_{i,j}$ is the $j$-th statistical feature extracted from the $i$-th time-domain phase of sample $x$, and $v_{m,n}$ is the $n$-th statistical feature extracted from the $m$-th value-domain segment of sample $x$.

*3.1.2 Effective feature selection*

To detect the degradation of the PM accurately, we need to select features that effectively represent the degradation. Over time, the degradation state of the PM will evolve into a fault state. In the meantime, the difference between the degradation feature and fault feature will gradually diminish and eventually disappear. Thus, we can take the fault feature as the indicator for measuring degradation. By extracting the statistical features of the power samples, a 64-dimensional feature set can be obtained. However, some features in this data set cannot effectively represent the fault state of the PM, so we need to select the features of the data set and obtain an effective feature set that can characterize the fault state. Compared to the normal state, some features of the fault state will deviate from the normal value, and the greater the degree of deviation, the more obvious is the discriminability of features from the fault. In this paper, we use the Fisher criterion [22] to analyze the fault state and normal state of the PM, and select the effective features that can represent and distinguish the fault state. The steps of feature selection are as follows:

(1) Calculate the intra-class variance and inter-class variance of statistical features between the fault state and normal state:

$$S_W^{(d)} = (m_F^{(d)} - m_N^{(d)})^2, \ d = 1, \cdots, 64 \tag{2}$$

$$S_B^{(d)} = (\sigma_F^{(d)})^2 + (\sigma_N^{(d)})^2, \ d = 1, \cdots, 64 \tag{3}$$

where $S_W^{(d)}$ and $S_B^{(d)}$ are the intra-class variance and inter-class variance of the $d$-dimensional feature, respectively. $m_F^{(d)}$ is the mean of the $d$-dimensional feature of the fault state. $m_N^{(d)}$ is the mean of the $d$-dimensional feature of the normal state. $\sigma_F^{(d)}$ is the standard deviation of the $d$-dimensional feature of the fault state. $\sigma_N^{(d)}$ is the standard deviation of the $d$-dimensional feature of the normal state.

(2) Calculate the criterion value between the fault state and normal state：

$$J^{(d)} = S_B^{(d)} / S_W^{(d)}, \ d = 1, \cdots, 64 \tag{4}$$

where $J^{(d)}$ is the criterion value of the $d$-dimensional feature between the fault state and normal state.

(3) For the 64 criterion values obtained, we take half of the maximum criterion value as the standard value, and select the feature whose criterion value is larger than the standard value.

(4) Calculate the correlation coefficient between the selected features of the fault state:

$$\rho_{pq} = \frac{\sum_i (fs_i^{(p)} - m_F^{(p)})(fs_i^{(q)} - m_F^{(q)})}{\sqrt{(\sigma_F^{(p)})^2}\sqrt{(\sigma_F^{(q)})^2}} \quad (5)$$

where $\rho_{pq}$ is the correlation coefficient between the $p$- and $q$-dimensional features of the fault state. $fs_i^{(p)}$ is the $p$-dimensional feature value of the $i$-th fault sample. The range of $\rho_{pq}$ is $(-1, 1)$. The larger the value of $|\rho_{pq}|$, the greater is the redundancy between the two features. If $|\rho_{pq}|$ is zero, it means that these two features are independent.

(5) For the correlation coefficient between the pairwise features, if the coefficient is greater than 0.95, then the feature with a higher criterion value is selected. Eventually all the remaining features are the selected effective features.

### 3.1.3 Feature set dimensionality reduction

By extracting and selecting the features of the power data, we can obtain the effective feature set that represents the operating state of the PM. For the SOM and SVM models, the use of a high-dimensional data set for learning will lead to overfitting problems, which affects the recognition accuracy [23]. Hence, we need to reduce the dimensionality of the feature set further to obtain the low-dimensional data set for model learning. Kernel principal component analysis (KPCA) is an improved method of PCA. Its core is to nonlinearly map the input vector to the high-dimensional feature space, and use PCA to calculate the principal component (PC) of the data in this feature space [24]. The linear PC obtained in the high-dimensional feature space is essentially the nonlinear PC in the original input space. For the characteristics of high dimensionality and non-linearity of the feature set, KPCA is used to extract its PCs in this work in order to reduce the feature set to the appropriate dimensionality and retain the original feature information to the maximum extent.

### 3.2 Degradation state mining

### 3.2.1 Self-organizing feature map

SOM is a self-organizing competitive neural network proposed by Teuvo Kohonen in the 1980s [25]. SOM consists of an input layer and competition layer, and its structure is shown in Fig. 7. On the left is the input layer, the number of neurons in this layer being $m$, which is the same as the dimension of the sample vector $X = (x_1, \cdots, x_m)$. On the right is the competitive layer, also known as the output layer, where the $n$ neurons of this layer are arranged in a two-dimensional array. The two layers of neurons are fully connected by variable weights, and the connection weight is $\omega_{ij}$ $(i = 1, \cdots, m, j = 1, \cdots, n)$. Thus, the neuron $j$ in the competitive layer has an $m$-dimensional weight vector $w_j$. Through the unsupervised learning mechanism, SOM makes the neurons in the competitive layer sensitive to the features of specific input vectors, and realizes that neurons act as recognizers of input vectors. After network learning, all input data is divided into different neuron regions of the competitive layer. The features of data within the region are similar, and the features of data between the regions are different, in order to realize the clustering analysis of the data.

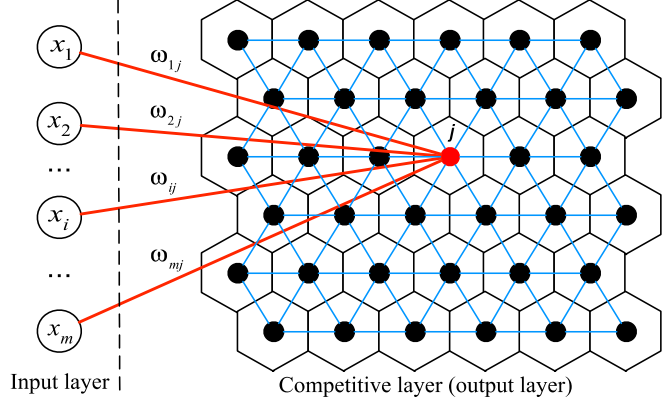

Fig. 7. Topological structure of the SOM network

The SOM learning algorithm is summarized as follows [26]:

(1) Set the learning iteration number $t = 0$. Initialize connection weights $\omega_{ij}$ with small random values. Set the initial topology neighborhood $N_0$. Set the initial learning rate $\eta_0$. Set the total number of iterations $T$.

(2) While iteration number $t$ is less than $T$, repeat Step 3–Step 6.

(3) Randomly select a vector $X(t)$ in the data set as the input vector.

(4) Calculate the Euclidean distance between the neurons of the competitive layer and the input vector:

$$d_j = \sqrt{\left|\sum_{i=1}^{m}(x_i(t) - \omega_{ij}(t))^2\right|}, j = 1, \cdots, n. \qquad (6)$$

Obtain the neuron $j^*$ that is the closest to the input vector and call it the winner neuron, which satisfies $d_{j^*} = \min_j(d_j)$.

(5) Update the weight of both the winner neuron $j^*$ and its neighborhood:

$$w_j(t+1) = w_j(t) + \eta(t)(X(t) - w_j(t)), \forall j \in N_{j^*} \qquad (7)$$

where $\eta(t) \in (0,1)$ is the learning rate function, which decreases exponentially with time until it tends to 0 to ensure the convergence of the learning process.

(6) Set $t = t + 1$, and if $t < T$, go back to Step 3, otherwise stop.

### 3.2.2 Power data for mining

The type of PM power data will directly affect the result of degradation state mining. To select the right type of power data, we specifically analyze the feature distribution of the 600 power samples acquired by the CSM. Fig. 8 shows the distribution of the first PC of these samples. It can be seen that the attribute values of

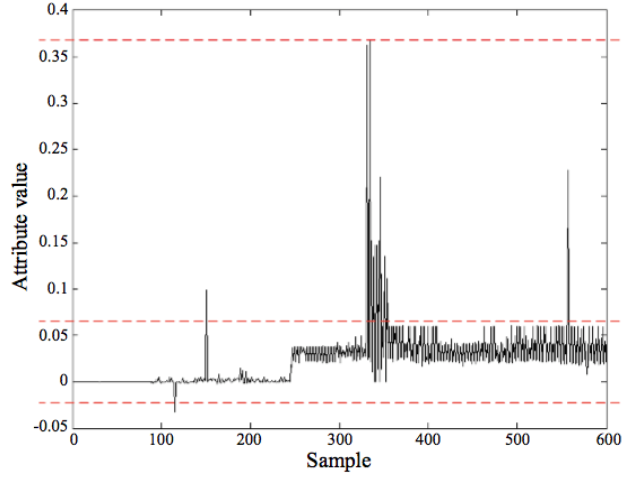
Fig. 8. First PC distribution of fault and non-fault samples

non-fault samples are distributed between 0 and 0.08, while the attribute values of fault samples are distributed between 0.08 and 0.37. Non-fault samples contain both normal state and latent state samples, where the attribute values of the normal samples are mostly 0, and a few slightly fluctuate around 0. The attribute values of the latent samples are mainly distributed between 0.02 and 0.08. We can find that the distributions of attribute values of normal and latent states are significantly different, proving that these two states can be effectively distinguished. Thus, we can use SOM to cluster non-fault-type power samples, mine the valid latent state samples, and remove normal state samples to obtain the degradation state samples of the PM.

### *3.2.3 Degradation state mining based on SOM*

Low-dimensional feature set $X_{nf} = \{x_i | i=1,\cdots,N\}$ can be obtained by performing feature processing on the collected nonfault samples, where $N$ is the number of samples. In this study, we used SOM to cluster the feature set multiple times and analyze each clustering result to obtain the degradation states of a PM with different characteristics. The steps of using SOM to cluster the feature set are described as follows:

(1) The number of neurons in the input layer of the SOM network are set to be the same as the dimension of the feature vector, and the competitive layer neurons are arranged in a two-dimensional (2D) array of $sn \times sn$. Then, feature set $X_{nf}$ is input for the first clustering. After clustering, the label number of the competitive layer neuron corresponding to each sample in $X_{nf}$ is counted to obtain clustering category set $clusA = \{a_{1,i} | i=1,\cdots,N\}$, where $a_{1,i}$ is the label number corresponding to the $i$th nonfault sample.

(2) The competitive layer neurons are arranged in a 2D array of $(sn + 1) \times (sn + 1)$. Then, feature set $X_{nf}$ is input for the second clustering, after which the clustering category is counted and set $clusB = \{b_{1,i} | i=1,\cdots,N\}$ is obtained.

(3) The competitive layer neurons are arranged in a 2D array of $(sn + 2) \times (sn + 2)$. Then, feature set $X_{nf}$ is input for the third clustering, after which the clustering category is counted and set $clusC = \{c_{1,i} | i=1,\cdots,N\}$ is obtained.

(4) *clusA*, *clusB*, and *clusC* are integrated to obtain the clustering sequence set $Seq_1 = \{(a_{1,i}, b_{1,i}, c_{1,i}) | i=1,\cdots,N\}$, where $(a_{1,i}, b_{1,i}, c_{1,i})$ is the sequence corresponding to the

*i*th nonfault samples.

To mine the degradation state of the PM, we propose a latent state analysis strategy. By analyzing the distribution of neurons in the competitive layer, the nonfault samples were selected, merged, and removed to obtain the dataset containing only degradation state samples. The steps for this strategy are described as follows.

(1) Count the label number of neurons in the *sn* × *sn* competitive layer, the number of clustering samples of which is not less than $\frac{M}{sn \times sn}$, and create neuron label set $nNum_1$. According to the first-dimensional data of $Seq_1$, select the samples corresponding to the label number of $nNum_1$ in $X_{nf}$. Create dataset $Sel_1 = \{x_i' | i = 1, \cdots, P\}$ and the corresponding clustering sequence set $Seq_2 = \{(a_{2,i}, b_{2,i}, c_{2,i}) | i = 1, \cdots, P\}$ by using the selected samples, where *P* is the number of selected samples.

(2) Count the label number of neurons in the (*sn* + 1) × (*sn* + 1) competitive layer, the number of clustering samples of which is not less than $\frac{M}{(sn+1) \times (sn+1)}$, and create neuron label set $nNum_2$. According to the second-dimensional data of $Seq_2$, select the samples corresponding to the label number of $nNum_2$ in $X_{nf}$. Create dataset $Sel_2 = \{x_i'' | i = 1, \cdots, Q\}$ and the corresponding clustering sequence set $Seq_3 = \{(a_{3,i}, b_{3,i}, c_{3,i}) | i = 1, \cdots, Q\}$ by using the selected samples, where *Q* is the number of selected samples.

(3) Count the label number of neurons in the (*sn* + 2) × (*sn* + 2) competitive layer, the number of clustering samples of which is not less than $\frac{M}{(sn+2) \times (sn+2)}$, and create neuron label set $nNum_3$. According to the third-dimensional data of $Seq_3$, select the samples corresponding to the label number of $nNum_3$ in $X_{nf}$. Create data set $Sel_3 = \{x_i''' | i = 1, \cdots, R\}$ and the corresponding clustering sequence set $Seq_4 = \{(a_{4,i}, b_{4,i}, c_{4,i}) | i = 1, \cdots, R\}$ by using the selected samples, where *R* is the number of selected samples.

(4) According to the distance distribution of the neurons in the (*sn* + 2) × (*sn* + 2) competitive layer, analyze the neighborhood of each neuron in $nNum_3$. Merge the neighboring neurons closet to them, and the labels are consistent with the neurons with the largest number of clustering samples. By referring to the merged neuron label numbers, modify the third-dimensional data of $Seq_4$ to update the corresponding clustering sequences of the $Sel_3$ samples and obtain $Seq_5 = \{(a_{5,i}, b_{5,i}, c_{5,i}) | i = 1, \cdots, R\}$.

(5) According to the distance distribution of the neurons in the (*sn* + 1) × (*sn* + 1) competitive layer, merge the neurons closet to neurons in $nNum_2$. By referring to the merged neuron label numbers, modify the second-dimensional data of $Seq_5$ to update the corresponding clustering sequences of the $Sel_3$ samples and obtain $Seq_6 = \{(a_{6,i}, b_{6,i}, c_{6,i}) | i = 1, \cdots, R\}$.

(6) According to $Seq_6$, sort the samples in $Sel_3$ to their respective clustering states. By analyzing the power curve of each state, find and remove the normal state samples. Finally, the degradation-state data set is obtained. Fig. 9 shows the overall flow of this proposed strategy.

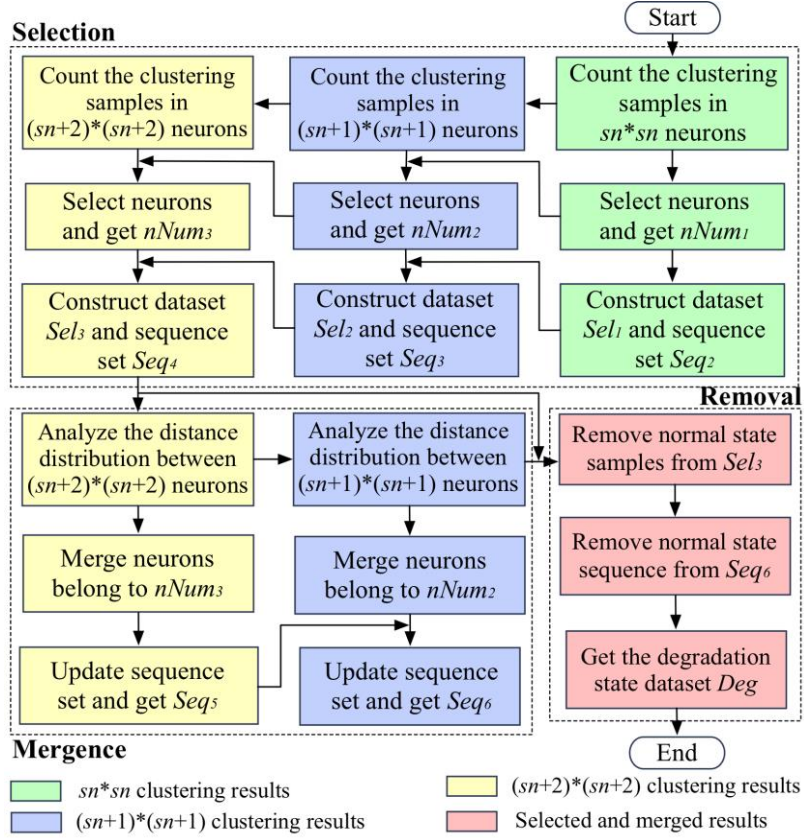
Fig. 9. Flowchart of the analysis strategy

## 3.3 Degradation-state identification

### 3.3.1 SVM

SVM, a machine learning method proposed by Vapnik and based on a statistical theory, is a common tool for pattern recognition. The core of SVM is to construct an optimal hyperplane with maximum margin, thus linearly separating the two classes of samples [27]. Let $S = \{(x_i, y_i) | x_i \in R^N, y_i \in \{-1,1\}, i = 1,\cdots,n\}$ be the training sample set, where $x_i$ is an input vector and $y_i$ is the label of $x_i$. The hyperplane function is given as

$$f(x) = \text{sgin}(w^T \cdot x + b). \tag{8}$$

To maximize the hyperplane margin, the following optimization problem should be solved:

$$\text{Minimize } \phi(w) = \left(\frac{1}{2} w^T w + C \sum_{i=1}^{n} \xi_i\right), \tag{9}$$

subject to

$$y_i(w^T \cdot \phi(x_i) + b) \geq 1 - \xi_i, \ i = 1,\cdots,n, \tag{10}$$

where $w$ is the weight coefficient vector of the hyperplane, $\phi(\cdot)$ is a mapping function, $\xi_i > 0$ is the slack variable, and $C$ is the penalty parameter used to control the degree of penalty for SVM misclassified samples. By using the Lagrange multiplier method and introducing the kernel function, the above-mentioned problem is equivalent to maximizing Eq. (11) under the constrained Eq. (12):

$$\text{Maximize } w(\alpha) = \sum_{i=1}^{n} \alpha_i - \frac{1}{2} \sum_{i,j=1}^{n} y_i y_j \alpha_i \alpha_j K(x_i, x_j) \tag{11}$$

$$\sum_{i=1}^{n} y_i \alpha_i = 0, \ 0 \leq \alpha_i \leq C, \ i=1,\cdots,n \qquad (12)$$

where $\alpha_i$ is a Lagrange multiplier and $K(x_i, x_j)$ is the kernel function representing inner product $\langle \phi^T(x_i) \cdot \phi(x_i) \rangle$. Therefore, the optimal hyperplane function takes the following form:

$$f(x) = \text{sgin}(\sum_{i=1}^{n} \alpha_i y_i K(x, x_i) + b). \qquad (13)$$

As shown in Fig. 10, the SVM nonlinearly maps the linearly inseparable training data in the input space to the high-dimensional feature space through kernel function $K$, making them linearly separable. In this study, we used the radial basis function (RBF) as the kernel function of the SVM.

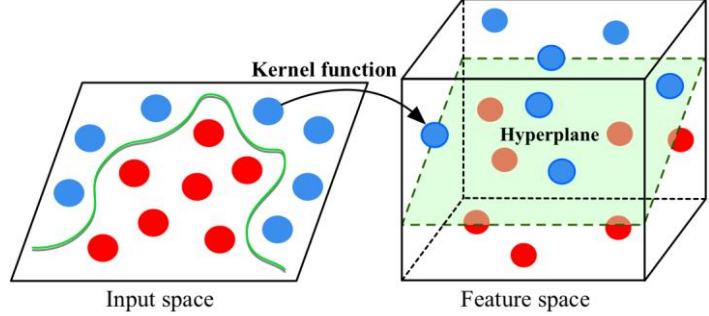

Fig. 10. Schematic of the kernel function

RBF is defined as follows:

$$K(x, x_i) = \exp(\| x - x_i \|^2 / 2\sigma^2), \ i=1,\cdots,n \qquad (14)$$

where $\sigma$ is a parameter that sets the "spread" of the kernel.

A multiclass SVM classifier is applied for identifying degradation because of the multiple state types existing in the PM. The multiclass SVM is mainly implemented by combining multiple binary SVMs. We used the "one-against-one" strategy to build the classifier [28]. For sample data of $k$ classes, we must build a $k(k-1)/2$ binary SVM. Each of these SVMs only considers two classes of samples, and all of the SVMs classify and vote on the input sample; the sample belongs to the class with the most votes.

### 3.3.2 PSO

For SVM, parameters $C$ and $\sigma$ have an impact on its classification performance. Thus, in the process of SVM training, the optimal parameters need to be selected to get high classification accuracy (CA). In this study, we used the PSO algorithm to select the optimal parameters of the SVM. The PSO algorithm is a population-based search algorithm, in which different particles have a fitness value determined by the objective function, and each particle is a point in the solution space with a certain velocity. The particles track the current optimal particle in the solution space and search for the optimal solution through iteration [c9]. Let the dimension of the target search space be $n$, the total number of particles be $m$, the position of particle $i$ be $X_i = (x_{i1}, \cdots, x_{in})$, and the velocity of particle $i$ be $V_i = (v_{i1}, \cdots, v_{in})$. So far, the best position searched by particle $i$ is $P_i = (p_{i1}, \cdots, p_{in})$, also known as local extremum. The best position searched by the whole particle swarm is $P_g = (p_{g1}, \cdots, p_{gn})$, also known as global extremum. In each iteration, particle $i$ updates the velocity and position as follows:

$$v_{id}^{k+1} = wv_{id}^k + c_1 r(p_{id} - x_{id}^k) + c_2 R(p_{gd} - x_{id}^k), \ i=1,\cdots,m, \ d=1,\cdots,n \tag{15}$$

$$x_{id}^{k+1} = x_{id}^k + v_{id}^{k+1}, \ i=1,\cdots,m, \ d=1,\cdots,n, \tag{16}$$

where $c_1$ and $c_2$ are acceleration constants, $r$ and $R$ are both random numbers in range [0,1] generated by the uniform probability distribution, and $v_{id}^k$ is the velocity of particle $i$ in the $k$th iteration; $v_{id}^k$ is restricted to the range $[-v_{i\max}, v_{i\max}]$ in which $v_{i\max}$ is a predefined boundary value. Further, $w$ is the inertia weight function used to balance the global and local search abilities of particles, as is defined as

$$w = w_{\max} - \frac{t}{t_{\max}}(w_{\max} - w_{\min}), \tag{17}$$

where $w_{\min}$ and $w_{\max}$ are the minimum and maximum values of $w$, respectively, $t$ is the current iteration number, and $t_{\max}$ is the maximum iteration number.

In this paper, the steps of using the PSO algorithm to optimize SVM parameters are described as follows:

(1) Randomly initialize the velocity and position of each particle ($C$ and $s$ parameters for SVM). Set the number of particles, maximum iteration number, inertia weights, acceleration constants, and search range.

(2) Build the SVM model based on the obtained parameters. Use the $k$-fold cross validation (K-CV) to train the model. Calculate the fitness of each particle in the population, i.e., average AC under K-CV.

(3) Update the optimal position of the individual particles and the global optimal position according to the fitness.

(4) Update the speed and position of each particle in the population.

(5) If the termination criterion is met, stop the iteration and obtain the SVM optimal parameters. Otherwise go back to Step 2. The termination criterion can be the maximum iteration number or the error precision of the fitness.

### 3.3.3 Degradation-state identification based on PSO–SVM

The essence of degradation-state identification is pattern recognition, and the degradation state of the PM can be specifically refined into different types of states. Therefore, we must build a multiclass classifier to identify the different degradation states of the PM. The process of multitype state identification based on PSO–SVM is shown in Fig. 11 and is divided into three steps:

(1) Data processing. The degradation state samples are categorized into training and test sets. Feature processing is performed on these sets to obtain the corresponding low-dimensional data sets *tr* and *te*, respectively.

(2) Training. The SVM is trained using *tr*, and the PSO algorithm is used to iteratively optimize the parameters of SVM until the termination criteria is met. Based on the "one-against-one" multiclass classification strategy, the classifier model is built by using the optimal parameters obtained.

(3) Test. *te* is used to test the CA of the model to prove the feasibility of employing it for identifying the degradation state of the PM.

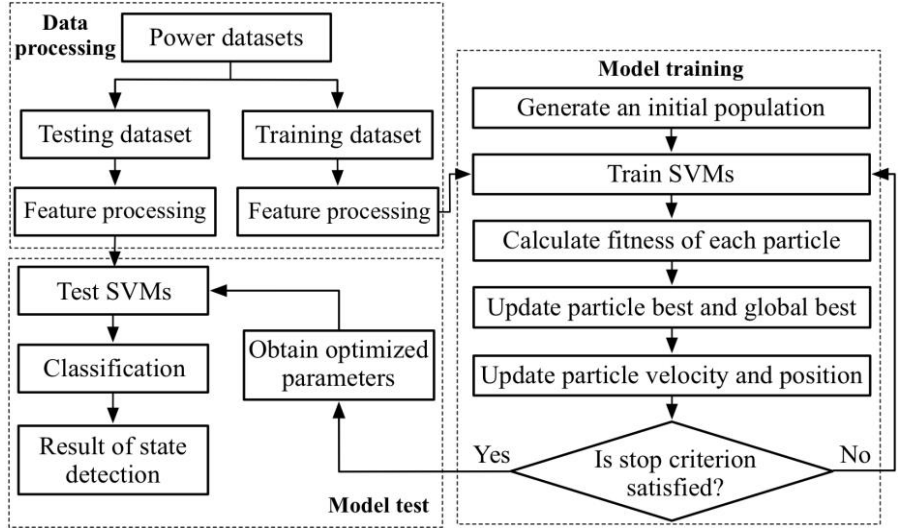
Fig. 11. Structural diagram of degradation identification based on PSO–SVM

## 4. Experimental results

In this study, we selected the 52-day power data set of the S700K PM corresponding to the No. 1 (J1, J2, J3), No. 2 (J1, J2, J3), No. 3 (J1), and No. 4 (J1), turnouts acquired by the CSM of Chenzhou West station, as the experimental data to verify the validity of the proposed method. This data set contains 20 normal samples, 1,000 nonfault samples, and 120 fault state samples.

**4.1 Typical faults of S700K PM**

Affected by a number of factors in relation to intensity of use, repair, maintenance, and environment, different types of faults occur on the PM. By investigating the actual working condition of the onsite equipment and reviewing relevant information, we summarized six typical faults of the S700K PM, as described in Table 1.

**Table 1**
Description of typical fault states

| Fault | Symptom | Cause | Location | Freq |
|---|---|---|---|---|
| F1 | Point machine experiences difficulty in unlocking the turnout rails | The tongue rail is too closely attached to the stock rail | Rail components of the turnout | 1 |
| F2 | Point machine experiences difficulty in switching the turnout rails | During the switch process, the friction among the rail components is too large | Rail components of the turnout | 1 |
| F3 | Point machine is idle during the switch process | The driving rod of the point machine is stuck | Point machine | 3 |
| F4 | Point machine is idle during the lock process | The detector rod of the point machine is blocked while moving | Point machine | 3 |
| F5 | The power values of the point machine in the slow-release area are very large | The diode of the outdoor rectifier stack is short-circuited | Indication circuit | 2 |
| F6 | The power values of the point machine in the slow-release area are all zero | The outdoor rectifier stack is open-circuited or the indoor control circuit is damaged | Indication circuit | 2 |

Values 1, 2, and 3 represent the frequency of fault occurrence.

Among them, *F1–F4* are mechanical faults, and *F5* and *F6* are circuit faults. In addition, *F1* and *F2* are caused by abnormalities of the rail components in the turnout system, while *F5* and *F6* are caused by control circuit problems of the turnout system. This indicates that a fault in the turnout system would directly cause the PM to fail. Fig. 12 shows the power curve of these six faults. By analyzing the curves, we summarize their characteristics, listed in Table 2, in the time-domain phase. For convenience, the four time-domain phases of start–release, switch, lock, and indication are represented as phases 1, 2, 3, and 4, respectively.

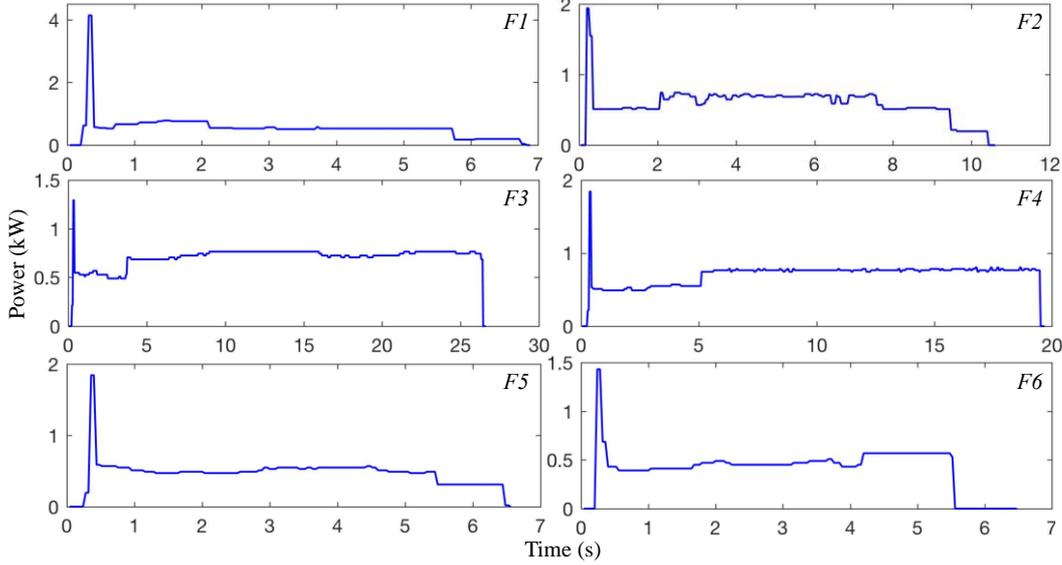

Fig. 12. Power curves of six typical faults

**Table 2**
Description of the fault characteristics in the time-domain phase

| Fault | Characteristic |
|---|---|
| *F1* | In phase 1, the power peak is overlarge and the overall power value is larger than normal condition |
| *F2* | In phase 2, the degree of power fluctuation is large |
| *F3* | In phase 2, the power value rises sharply and remains constant until it reaches the time limit and then directly drops to zero |
| *F4* | In phase 3, the power value rises sharply and fluctuates abnormally until it reaches the time limit and then directly drops to zero |
| *F5* | In phase 4, the power value is approximately twice than that of the normal condition |
| *F6* | In the later part of phase 4, the power value drops and remains at zero |

Further, we projected the six power curves onto the value domain. Fig. 13 shows that the power distributions of these fault types have significant differences. The power peak of *F1* is large, that is approximately 4.2 kW, which is caused by the high power of the PM to release the turnout during phase 1. The power distribution of *F2* is relatively discrete, which is caused by the large power fluctuation of the PM switching the turnout during phase 2. *F3* and *F4* have more power points, resulting from the long idle time of the PM in phases 2 and 3, respectively. In addition, the power values of these two faults are mostly distributed between 0.5 and 1 kW, indicating that the operating power of the PM is high in these two states. The power value of *F5* has a specific distribution between 0 and 0.5 kW, which is caused by the abnormal power change in the slow-release region during phase 4. *F6* has more power

"zero value", which is caused by the power values of the slow-release region are all zero during the phase 4.

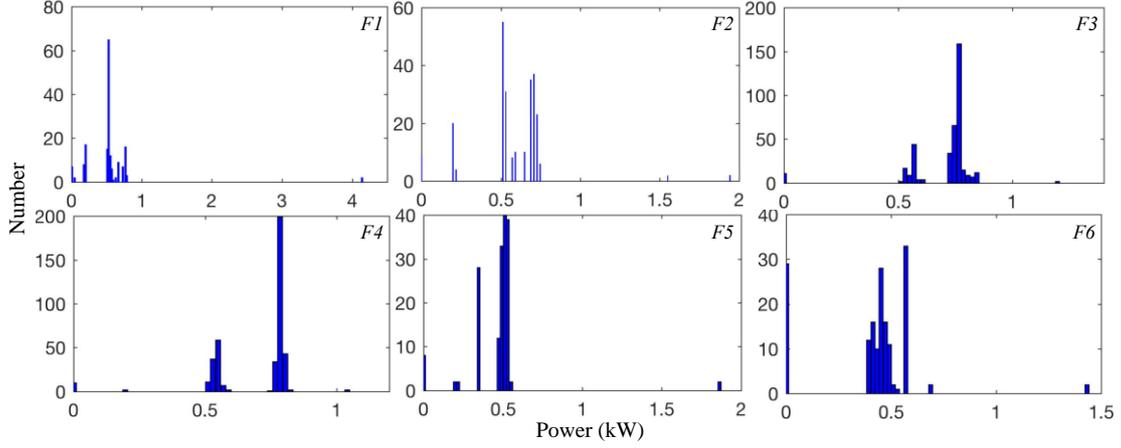

Fig. 13. Value domain projection of six typical faults

**4.2 Power data feature processing results**

We selected 20 normal samples (*N*) and 20 samples of each fault state to form a standard data set containing 140 samples. By performing feature processing on this data set, the effective features of the PM and the setting of KPCA dimensionality reduction parameters can be obtained. Further, with reference to the above-mentioned process, we preformed feature processing on the nonfault data set used for degradation state mining.

By extracting the statistical parameters of the standard data set from the time and value domains, the 64-dimensional feature set was obtained. Then, we calculated and compared the criterion values and correlation coefficients of this feature set, and selected the effective features, that is, $t_{1,3}$, $t_{2,1}$, $t_{2,2}$, $t_{2,8}$, $t_{4,3}$, $t_{4,5}$, $t_{4,8}$, $t_{4,9}$, $v_{1,2}$, $v_{1,7}$, and $v_{3,3}$, corresponding to the 4th, 11th, 12th, 18th, 33th, 35th, 38th, 39th, 42nd, 47th, and 59th dimension features of the set. These feature data were used to form a new 11-dimensional feature set. We normalized this data set and obtained the distribution of attribute values of the samples in each feature dimension, as shown in Fig. 14. These 11 selected features can effectively distinguish the different states of the sample data. Furthermore, we used KPCA to reduce the dimensionality of the feature set.

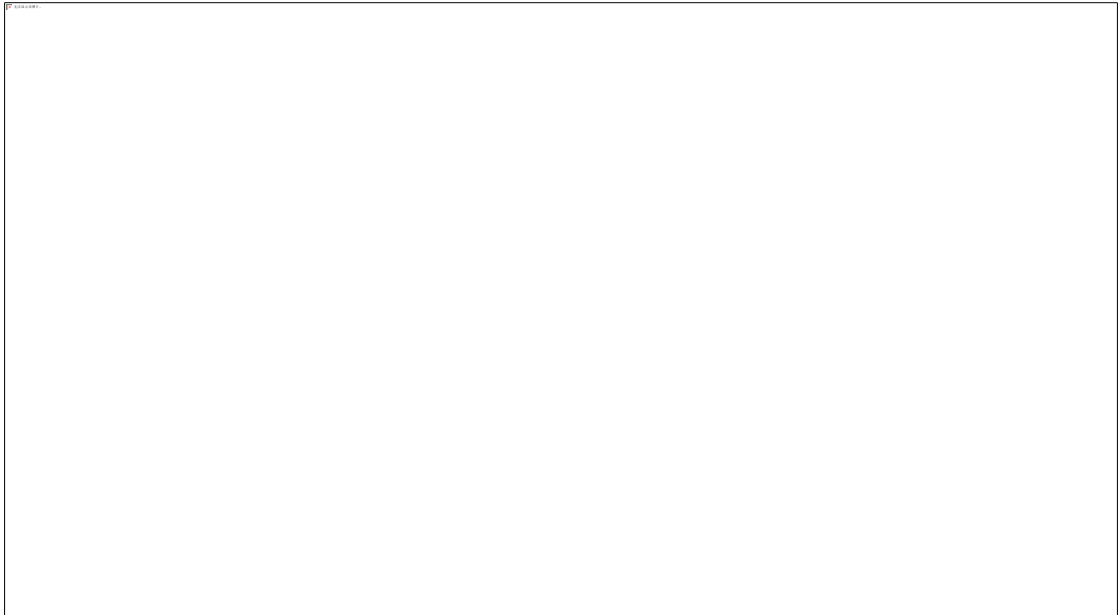
Fig. 14. Distribution of normal and fault samples in the effective feature dimensions

Here, the kernel function of KPCA was set as RBF and radial basis parameter $s = 2$. The number of PCs was set to 3–9. By comparing these 7-dimensionality reduction results, we find that when the number of PCs was set to 6, i.e., the dimensionality of the data set was reduced to 6 dimensions, the distinction among the different states was the best, and more than 99% of the original information could be retained. The distribution of attribute values of the samples in different PCs is shown in Fig. 15. Feature processing was performed on the nonfault data set consisting of 1,000 samples in the same manner as the above-mentioned procedure, and the 6-dimensional feature set could be obtained. Fig. 16 shows the distribution of these nonfault samples in different PCs. As shown, the distribution of the samples differs in all the six dimensions, and the samples can be clearly distinguished. In this study, we used this data set for degradation mining to obtain multiple degradation-state data with different characteristics of the PM.

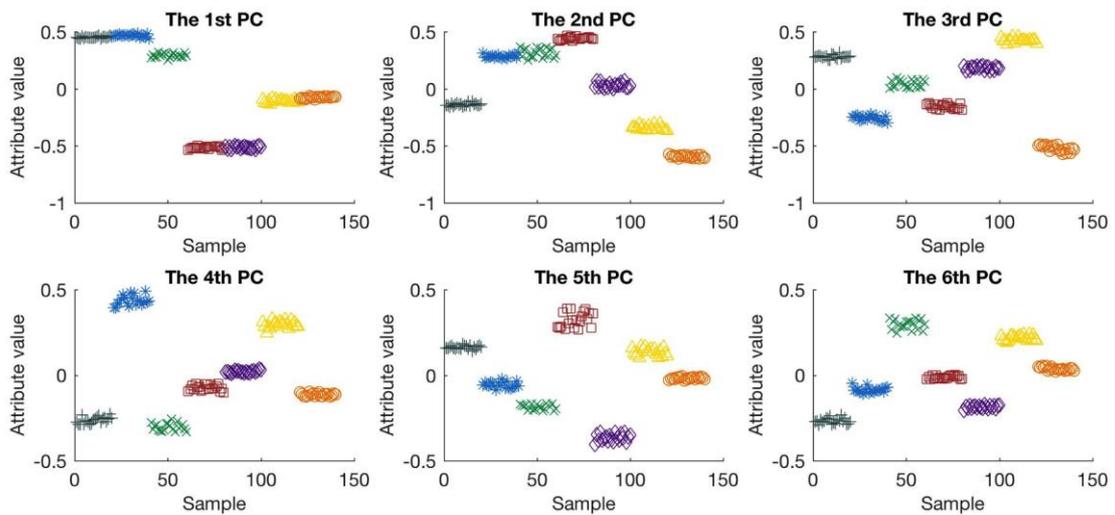
Fig. 15. Distribution of normal and fault samples in the six PCs

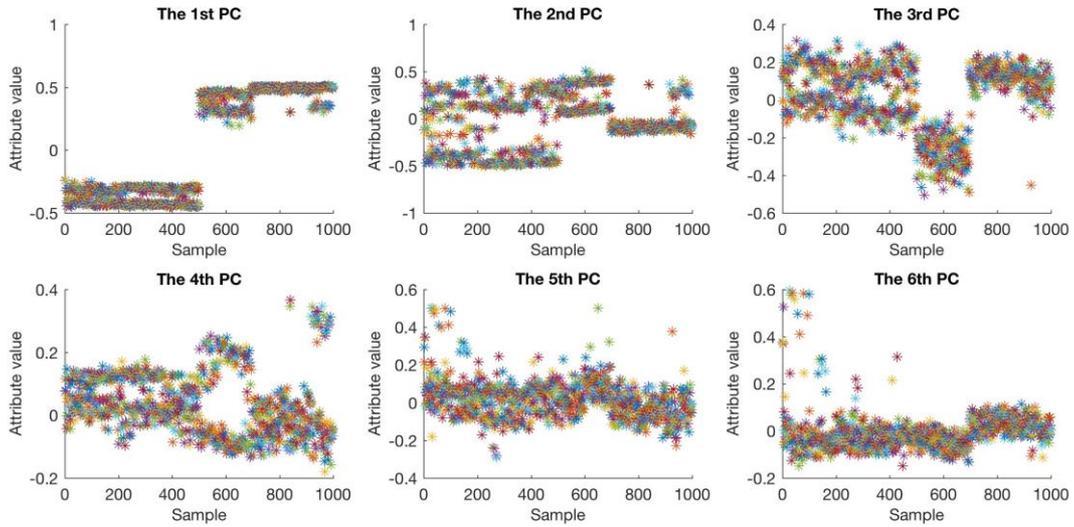

Fig. 16. Distribution of nonfault samples in the six PCs

### 4.3 Degradation-mining results

In this study, we used the SOM neural network toolkit developed by MathWork Inc., to conduct clustering analysis on the 6-dimensional non-fault data set, and used the proposed analysis strategy to select, merge, and remove the clustering results to mine the degradation-state data of the PM. The competitive layer neurons were arranged in 2D arrays of 4 × 4, 5 × 5, and 6 × 6, respectively. The connection weights of SOM were randomly initialized, and the neuron neighborhood shrinks in the shape of a hexagonal grid. The number of iterations and learning epochs was 1000 and 20, respectively. The loss function was a mean square error. After setting the SOM parameters, we conducted clustering analysis on the data set and obtained the clustering results. The distributions of the samples after clustering are shown in Fig. 17, where the hexagon lattice represents the neuron in the competitive layer and the number in the lattice represents the sample size clustered by the neurons. The clustering results show that with the increase of the number of neurons in the competitive layer, the distribution of samples becomes dispersed, complicating the analysis of the clustering results. Thus, we need to set the number of neurons reasonably and make a trade-off between the clustering effect and complexity. Fig. 18 shows the distance distribution of neurons in the competitive layer after clustering, where the connection band between adjacent neurons is used to measure the distance. The lighter color of the connection band represents a smaller distance and the darker color represents a larger distance.

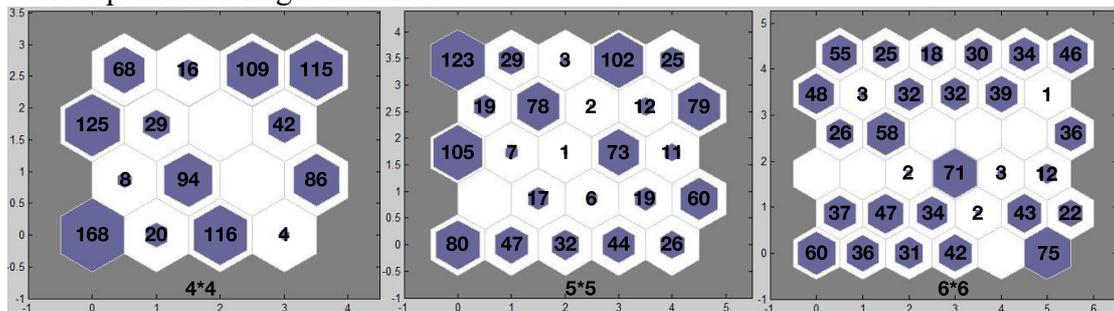

Fig. 17. Distribution of clustering samples in competitive layer neurons

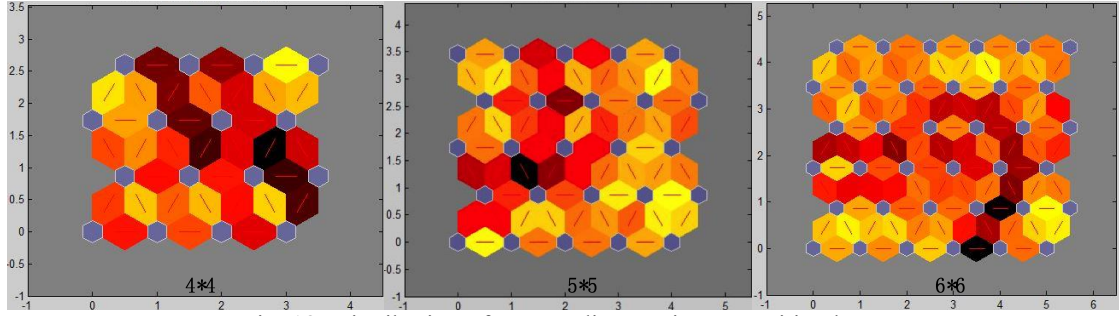
Fig. 18. Distribution of neuron distance in competitive layer

By using the analysis strategy to select, merge, and remove SOM clustering results, we obtained six degradation states of the PM: *D1–D6*. Table 3 shows the clustering sequence and sample size of each state. The total number of these degradation samples is 386, accounting for 38.6% of the nonfault data set. This indicates that most of the samples in the data set are not divided into degradation states. These undivided samples include normal and intermediate state samples. During the transition from normal to degradation state, the PM experiences several intermediate states. The difference between the intermediate and normal states is blurred, and compared with the degradation state excavated, the intermediate state is not typical to reflect the degradation of the PM. Therefore, they are not divided into degradation states. In the process of degradation mining, a data set with large capacity and various sample types should be used to obtain states that can represent the degradation characteristics of the PM.

**Table 3**
Statistical information of six degradation states

| Clustering sequence | *D1* | *D2* | *D3* | *D4* | *D5* | *D6* |
|---|---|---|---|---|---|---|
| SOM 4 × 4 | 1 | 6 | 8 | 9 | 3 | 15 |
| SOM 5 × 5 | 21 | 20 | 17 | 10 | 24 | 11 |
| SOM 6 × 6 | 7 | 20 | 16 | 29 | 31 | 1 |
| **Sample number** | 73 | 58 | 71 | 69 | 55 | 60 |

To determine whether these degradation states are valid, their characteristics must be specifically analyzed. Fig. 19 shows the power curves of the six degradation states. The power values of *D1*, *D2*, and *D3* are normally distributed in phase 1, 3, and 4, respectively. However, in phase 2, compared with *N*, the power value and fluctuation degree of the three states increase successively. The characteristics of *D1*, *D2*, and *D3* were determined to be similar to those of *F2*. This may be because the PM is subject to abnormal resistance in the process of switching the tongue rail. The power values of *D4* and *D5* are normally distributed in phases 2 and 4, respectively. However, in phase 1, the power peak of the two states is lower than that of *N*. Moreover, in phase 3, compared with *N*, the power value and fluctuation degree of *D4* and *D5* increase successively; these characteristics are consistent with those of *F4*. This may be because the detector rob moves abnormally in the process of the PM locking the tongue rail. The overall power value of *D6* is lower than that of *N*, while there is no abnormality in its fluctuation degree. Compared with the other five states, the degradation characteristic of this state is not obvious. By viewing the field maintenance log, we believe that this could be because the debugging of the PM is imperfect before operation or after maintenance.

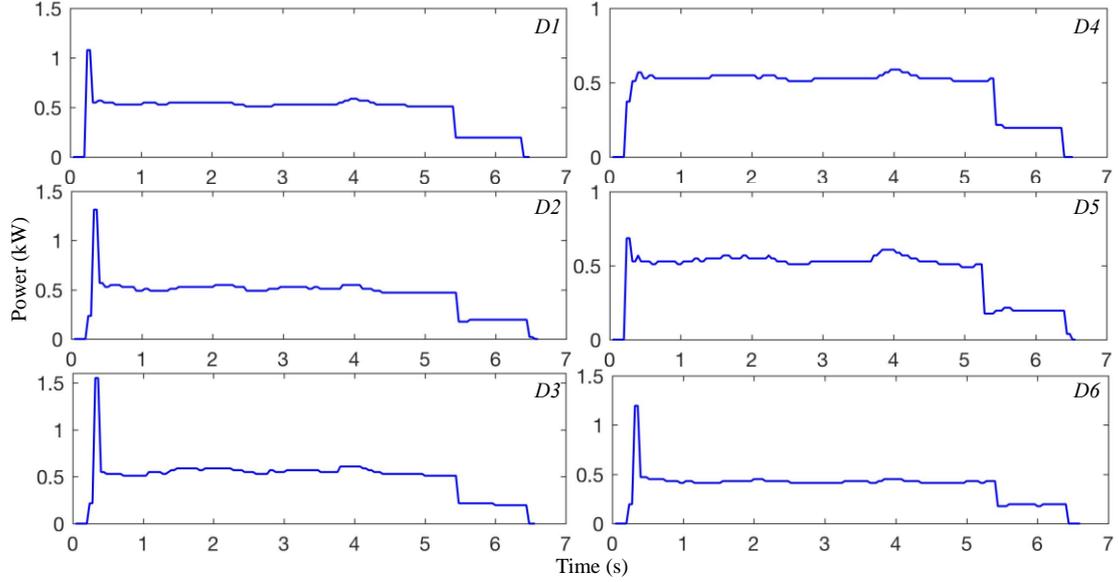

Fig. 19. Power curves of six degradation states

## 4.4 Degradation-identification results

Through the clustering analysis of nonfault data, six degradation states of the PM were mined and their corresponding samples were obtained. For these states, we created a degradation data set with 330 samples, in which the number of samples in each state is 55. Each state sample of the data set was randomly divided according to the ratio of 4:1, which was used to create the training and test sets. We performed feature processing on these two data sets and obtained the 6-dimensional input sets *tr* and *te*. Then, we set the parameters of the PSO–SVM as follows: inertia weights $w_{min}$ and $w_{max}$ of the PSO were 0.4 and 0.9, respectively; number of particles was 20; maximum iteration number was 200; and acceleration constants $c_1$ = 1.5 and $c_2$ = 1.7. The search ranges of the SVM parameters *C* and $\sigma$ were [0,100] and [0,1000], respectively, and the two parameters were randomly initialized. The 5-CV is used to evaluate the performance of SVM parameters. After setting parameters, we used *tr* and *te* to train and test the PSO–SVM, respectively. Five simulation experiments were conducted and the average CA of the model was considered as the final experimental result. Fig. 20 (a) shows the fitness curve of PSO searching for optimal parameters in one of the experiments, where the best fitness is 100% and the average fitness is approximately 96%. The optimal parameter (*C*, $\sigma$) of the SVM obtained in this experiment was (4.894, 20.954). Fig. 20 (b) shows the classification results of the above-mentioned experiment. In this experiment, the classifier mistakenly identified the two samples of *D6* as those of *D2* and *D3* states. As the degradation characteristics of *D6* are not obvious, and its distribution of power values in phases 1, 2, and 4 is similar to that of *D2* and *D3*, the model fails to accurately identify the samples of *D6*.

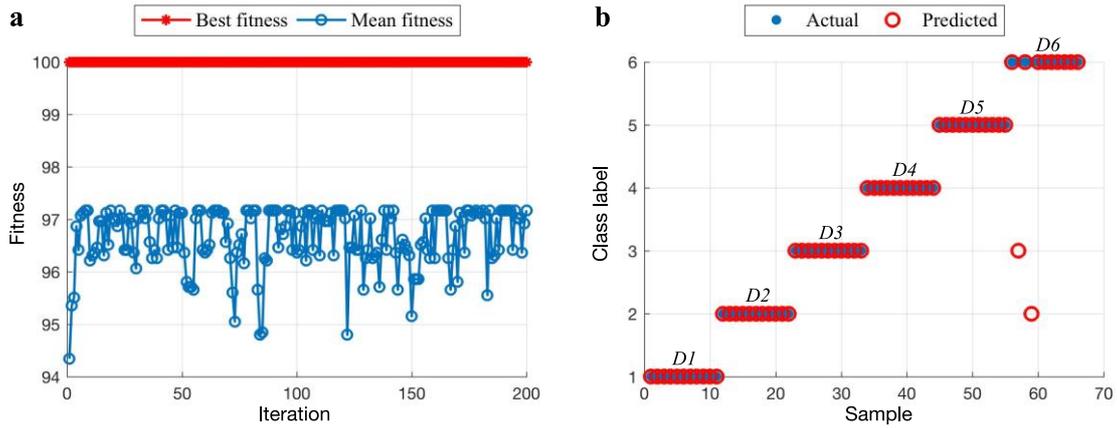
Fig. 20. PSO–SVM training and test: (a) fitness curve of PSO, (b) classification results.

The experimental results among different methods are listed in Table 4. As shown, the average CA of the proposed model is superior to those of others. Compared with the method based on PCA, the proposed method uses KPCA to extract the nonlinear information of the PM power data, and retains more original data for learning and classification. Compared with the method without the PSO algorithm, the proposed method uses the optimal parameters to build the model to improve the classification performance. The average CA of the last two methods is low, proving the necessity of further dimensionality reduction for the power feature data.

**Table 4**
Experimental results among different methods

| State | KPCA + PSO–SVM | PCA + PSO–SVM | KPCA + SVM | PSO–SVM | SVM |
|---|---|---|---|---|---|
| $D1$ (%) | 100 | 90.91 | 95.46 | 90.91 | 72.73 |
| $D2$ (%) | 100 | 100 | 100 | 100 | 81.82 |
| $D3$ (%) | 100 | 100 | 100 | 90.91 | 90.91 |
| $D4$ (%) | 100 | 81.82 | 86.37 | 81.82 | 90.91 |
| $D5$ (%) | 100 | 100 | 100 | 100 | 81.82 |
| $D6$ (%) | 86.37 | 81.82 | 72.73 | 72.73 | 81.82 |
| **CA** (%) | 97.73 | 92.42 | 92.43 | 89.39 | 80 |

## 5. Conclusions

Based on SOM and SVM, this paper proposed a degradation-detection method, which can detect the degradation signs of a PM before failure occurs so as to provide guarantee for its maintenance. The two major contributions of this study are as follows: 1) the mining of the degradation states of a PM with different characteristics and 2) the development of a classifier for accurately identifying the degradation state of the PM. The effectiveness of the proposed method was verified using the field-monitoring data sets. The results of the clustering analysis showed that the proposed method can mine multiple valid degradation states of the PM. The classifier can identify the degradation state of the PM with 96.97% accuracy. The study proved that the proposed method provides an effective means to detect the degradation of the PM for grasping its operation.

# Reference

[1] Tavner P J. Review of condition monitoring of rotating electrical machines[J]. Electric Power Applications Iet, 2008, 2(4):215-247.

[2] Márquez F P G, Roberts C, Tobias A M. Railway point mechanisms: Condition monitoring and fault detection[J]. Proceedings of the Institution of Mechanical Engineers Part F Journal of Rail & Rapid Transit, 2010, 224(1):35-44.

[3] Lee J, Choi H, Park D, et al. Fault Detection and Diagnosis of Railway Point Machines by Sound Analysis[J]. Sensors, 2016, 16(4):549.

[4] Vileiniskis M, Remenyte-Prescott R, Rama D. A fault detection method for railway point systems[J]. Proceedings of the Institution of Mechanical Engineers Part F Journal of Rail & Rapid Transit, 2016, 230(3):18.

[5] Aguiar E P D, Nogueira F M D A, Amaral R P F, et al. EANN 2014: a fuzzy logic system trained by conjugate gradient methods for fault classification in a switch machine[J]. Neural Computing & Applications, 2016, 27(5):1175-1189.

[6] Yilboga H, Ömer Faruk Eker, Güçlü A, et al. Failure prediction on railway turnouts using time delay neural networks[C]// IEEE International Conference on Computational Intelligence for Measurement Systems and Applications. IEEE, 2010:134-137.

[7] Böhm T. Accuracy Improvement of Condition Diagnosis of Railway Switches via External Data Integration[C]// European Workshop on Structural Health Monitoring. DLR, 2012:1550-1558.

[8] Kuang F, Xu W, Zhang S. A novel hybrid KPCA and SVM with GA model for intrusion detection[J]. Applied Soft Computing Journal, 2014, 18(C):178-184.

[9] Asada T, Roberts C, Koseki T. An algorithm for improved performance of railway condition monitoring equipment: Alternating-current point machine case study[J]. Tramsportation Research Part C Emerging Technologies, 2013, 30(5):81-92.

[10] Tran V T, Hong T P, Yang B S, et al. Machine performance degradation assessment and remaining useful life prediction using proportional hazard model and support vector machine[J]. Mechanical Systems & Signal Processing, 2012, 32(4):320-330.

[11] Guo L, Li N, Jia F, et al. A recurrent neural network based health indicator for remaining useful life prediction of bearings[J]. Neurocomputing, 2017, 240(C):98-109.

[12] Benkedjouh T, Medjaher K, Zerhouni N, et al. Health assessment and life prediction of cutting tools based on support vector regression[J]. Journal of Intelligent Manufacturing, 2015, 26(2):213-223.

[13] Tobon-Mejia D A, Medjaher K, Zerhouni N, et al. A Data-Driven Failure Prognostics Method Based on Mixture of Gaussians Hidden Markov Models[J]. IEEE Transactions on Reliability, 2012, 61(2):491-503.

[14] Widodo A, Yang B S. Machine health prognostics using survival probability and support vector machine[J]. Expert Systems with Applications, 2011, 38(7):8430-8437.

[15] Jiang H, Chen J, Dong G. Hidden Markov model and nuisance attribute projection based bearing performance degradation assessment[J]. Mechanical Systems & Signal Processing, 2016, s 72–73:184-205.

[16] An D, Kim N H, Choi J H. Practical options for selecting data-driven or physics-based prognostics algorithms with reviews[J]. Reliability Engineering & System Safety, 2015, 133:223-236.

[17] Yu J. A hybrid feature selection scheme and self-organizing map model for machine health assessment[J]. Applied Soft Computing Journal, 2011, 11(5):4041-4054.

[18] Germen E, Başaran M, Fidan M. Sound based induction motor fault diagnosis using Kohonen self-organizing map[J]. Mechanical Systems & Signal Processing, 2014, 46(1):45-58..

[19] Hu J, Zhang L, Liang W. Dynamic degradation observer for bearing fault by MTS–SOM system[J]. Mechanical Systems & Signal Processing, 2013, 36(2):385-400.

[20] Márquez F P G, Muñoz J M C. A pattern recognition and data analysis method for maintenance management[J]. International Journal of Systems Science, 2012, 43(6): 1014-1028.

[21] Li Y, Liang X, Lin J, et al. Train axle bearing fault detection using a feature selection scheme based multi-scale morphological filter[J]. Mechanical Systems & Signal Processing, 2018, 101:435-448.

[22] Yin L, Ge Y, Xiao K, et al. Feature selection for high-dimensional imbalanced data[J]. Neurocomputing, 2013, 105: 3-11.

[23] Kuo F Y, Sloan I H. Lifting the curse of dimensionality[J]. Notices of the AMS, 2005, 52(11): 1320-1328.

[24] Shao R, Hu W, Wang Y, et al. The fault feature extraction and classification of gear using principal component analysis and kernel principal component analysis based on the wavelet packet transform[J]. Measurement, 2014, 54(6):118-132.

[25] Ritter H, Martinetz T, Schulten K, et al. Neural computation and self-organizing maps: an introduction[M]. Reading, MA: Addison-Wesley, 1992.

[26] Kohonen T. The self-organizing map[J]. Neurocomputing, 1998, 21(1-3): 1-6.

[27] Adankon M M, Cheriet M. Support Vector Machine[J]. Computer Science, 2009, 1(4):1-28.

[28] Huang C L, Dun J F. A distributed PSO–SVM hybrid system with feature selection and parameter optimization[J]. Applied Soft Computing Journal, 2008, 8(4):1381-1391.